# The Importance of the Digital Object Identifier (DOI) in Enhancing the Credibility of Scientific Research: An Analytical Data Study

Ahmed Shaker Alalaq

Iraq/ University of Kufa

Ahmed.alallaq@uokufa.edu.iq

**Abstract**

This study aims to analyze the vital role played by the Digital Object Identifier (DOI) in enhancing the credibility and reliability of scientific research in the digital age. Through an analytical study of DOI usage data derived from international scientific publishing institutions, the extent of its adoption and its recognition as a global standard for encoding research and academic sources was highlighted.

The results showed that the number of scientific records registered using DOIs exceeded 167 million, with more than 30,000 DOI prefixes distributed across over 150 countries, reflecting the significant growth in its use by academic and research institutions. Additionally, more than 3.2 billion monthly DOI resolutions were recorded, indicating the increasing reliance on them for accessing resources.

The study also included an analysis of the content types registered with DOIs, showing that scientific articles constituted the majority at 71%, followed by books and conference papers. A notable finding was that 95% of citations linked to DOIs are now openly available, contributing to greater transparency and scientific verifiability.

The study concluded that the DOI is not merely an organizational tool but a central element in the structure of modern scientific publishing. It contributes to improving research quality, facilitating verification, and ensuring continued accessibility. The study recommended the broader adoption of DOIs, especially in emerging scientific communities, to achieve greater integration in the global research information infrastructure.

Keywords: Digital Object Identifier, Data, Standards, Systems, Journals, Indexing

## 1. Introduction

Scientific research in the digital age is witnessing rapid developments in publishing and documentation tools, and access to and verification of information has become a critical factor in ensuring quality and credibility. Among these tools, the Digital Object Identifier (DOI) stands out as a globally standardized system used to uniquely and permanently identify electronic academic resources. Thousands of journals and scientific institutions have adopted it as a means to ensure continued access to research content, regardless of changes in URLs or publishing platforms.

The importance of this topic stems from the DOI's direct impact on enhancing the credibility of scientific research by supporting transparency, facilitating the tracking and verification of sources, and enabling the construction of a reliable knowledge base that can be precisely referenced. It is also an effective tool for supporting the open access movement and integration between publishing and citation systems.

However, the actual use of the DOI is not without challenges. The most notable are disparities in its adoption across institutions and countries, the lack of standardized usage practices in some academic communities, and limited awareness of its true role in improving scientific quality. Moreover, a knowledge gap exists concerning the large-scale analysis of DOI usage data to assess its effectiveness and global reach.

This study is limited to analyzing data published by official bodies such as Crossref up to the year 2025, focusing on quantitative indicators related to DOI usage (e.g., number of records, geographical distribution, content types, volume of open citations, and more). Qualitative assessments of researchers' or institutions' experiences are excluded, keeping the scope within the realm of digital data analysis.

## 2. Historical Background

The concept of the Digital Object Identifier (DOI) system was introduced by the International DOI Foundation (IDF) in the year 2000, aiming to address the challenges associated with identifying and accessing digital objects, especially in academic and professional environments (Mondal, 2023). Before the creation of the DOI, the proliferation of digital content was hindered by the lack of standardized identification methods, making it difficult for users to consistently locate and cite research materials. The groundwork for the DOI system was laid in the late 1990s when the IDF was established in 1998. This initiative brought together a coalition of stakeholders, including publishers, libraries, and technology providers, to create a reliable and persistent linking system for digital content (Radadiya, 2025).

The DOI system was inspired by the Handle System, a decentralized protocol designed to assign unique identifiers to digital objects. It aimed to enhance content accessibility by establishing a centralized registration authority responsible for managing DOIs (Eduindex, 2024).

Since its inception, the DOI system has experienced significant growth. The number of registered DOIs increased from 50 million in 2011 to an expected 391 million by 2025, reflecting rising demand for digital content accessibility. This rapid expansion was accompanied by a rise in the number of participating organizations, growing from 4,000 members in 2011 to 9,500 by 2013, although the current exact number of members remains unclear due to the federal structure of the system (hub.tghn, n.d.).

The introduction of the DOI system has fundamentally transformed the landscape of scholarly communication, enabling researchers and the general public to reliably find, access, and cite digital objects—thereby facilitating the dissemination of knowledge in an increasingly digital world.

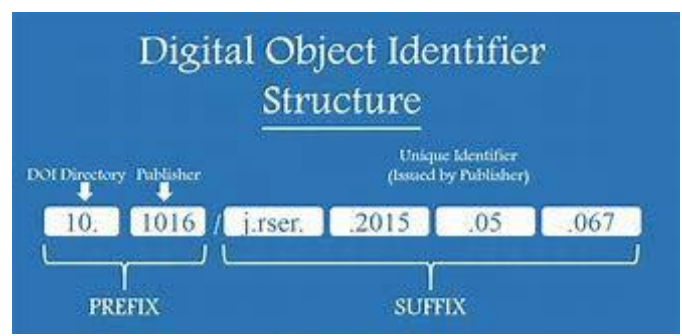

Figure (1) showing the structure of a digital identifier

DOI Structure

A DOI is typically structured as follows: Prefix/Suffix

- Prefix: A unique code assigned to the publisher, journal, or organization registering the Digital Object Identifier (DOI). For example, "10.1000" is a commonly used prefix for a specific publisher.
- Suffix: Assigned by the publishing entity to uniquely identify the specific item within their catalog. This suffix can consist of a combination of letters, numbers, or both.

An example of a DOI might look like this: 10.1038/s41586-019-1666-5 (Eduindex, 2024)

## 3. What is a DOI?

The Digital Object Identifier (DOI) is one of the most essential components of the modern scholarly publishing infrastructure. It provides a reliable and permanent method for identifying and tracking scientific content within an ever-changing digital environment. The DOI plays a critical role in enhancing the dissemination of scientific research, facilitating access, and ensuring the persistence of academic references—all of which contribute to the global advancement of scientific inquiry.

Recent research indicates that DOI facilitates an average of 1.1 billion monthly accesses to scientific content, representing 95% of all global digital identifier activity, underscoring its immense significance in the circulation and dissemination of scholarly knowledge (Isjem, n.d.).

The DOI is a unique and persistent alphanumeric identifier used to identify digital objects such as articles, books, and datasets (Editorial, 2017; Laakso & Hartgerink, 2022). It plays a vital role in digital content management and accessibility, especially within academic and research environments (Paskin, 1999).

A DOI consists of a unique alphanumeric string used to identify and provide a persistent link to digital objects, including research articles, datasets, and other types of scholarly content. It is a widely adopted standard used to direct readers to digital information consistently and permanently. The system was developed in response to the dynamic and fast-evolving nature of the digital landscape, where information can easily be lost as URLs change and digital objects move across the web. A key advantage of the DOI is its persistence—it remains the same even if the location of the digital object changes, making it a stable and permanent link to scientific content.

The DOI system is managed by the International DOI Foundation (IDF), which consists of registration agencies such as Crossref and DataCite. These agencies facilitate the assignment of DOIs to informational objects of academic and scientific value—such as journal articles, datasets, books, and book chapters. The DOI registration process includes assigning a unique alphanumeric string to the digital object, along with metadata that describes it—such as the title, authors, and publication date. Once a DOI is assigned to a scientific article, it remains unchanged even if the article's online location is altered. Instead, the core metadata supporting the DOI is updated to reflect the article's new URL.

### 4. Difference Between DOI and Regular URLs

The Digital Object Identifier (DOI) ensures stability and continuity in accessing academic resources, whereas regular links such as URLs may change or be removed over time. For this reason, using DOIs in scholarly work is preferred to guarantee permanent access to content. The table below illustrates the key differences between the two: (Davidson, L. A., & Douglas, 1998; Jasmin, 2023; Peukert, C., & Windisch, 2024).

Table No. 1 shows the difference between a digital ID and traditional links.

| Criterion | DOI (Digital Object Identifier) | Traditional URL |
|---|---|---|
| Persistence | Stable and unchanging, even if content is moved or the site structure is modified; centrally managed by the International DOI Foundation. | Prone to change or link rot when domains or folder structures are modified. |
| Timestamping | Officially timestamped in peer-reviewed databases (e.g., Crossref), supporting legal claims and scholarly citations. | Not officially recorded in any central registry; relies on local server timestamps if available. |
| Metadata | Includes structured metadata (authors, title, journal, publication year, publisher, license | Does not contain embedded metadata except for editable |

| | type, subject classification, etc.). | HTML meta tags. |
|---|---|---|
| IP Rights Protection | Directly linked to the publisher or rights holder; license (e.g., Creative Commons) can be embedded during registration. | May be redirected or hosted without copyright control; harder to verify the original source. |
| Scholarly Integration | Internationally recognized by academic journals, search engines, digital libraries, and citation indexes; increases credibility and citation ease. | Not formally recognized by scholarly indexing systems; may be excluded from databases. |
| Link Rot Resistance | Internal update mechanism to redirect DOI to new content location if the URL changes. | If the URL or page is deleted, the link becomes invalid and is not auto-redirected. |
| Retrievability | Content can be retrieved via global DOI resolvers (e.g., doi.org), regardless of hosting location. | Requires knowledge of the exact domain and path; no central service to retrieve lost links. |
| International Adoption | Widely adopted in academic publishing, official research releases, and institutional repositories. | Adoption varies by site; no unified managing authority. |

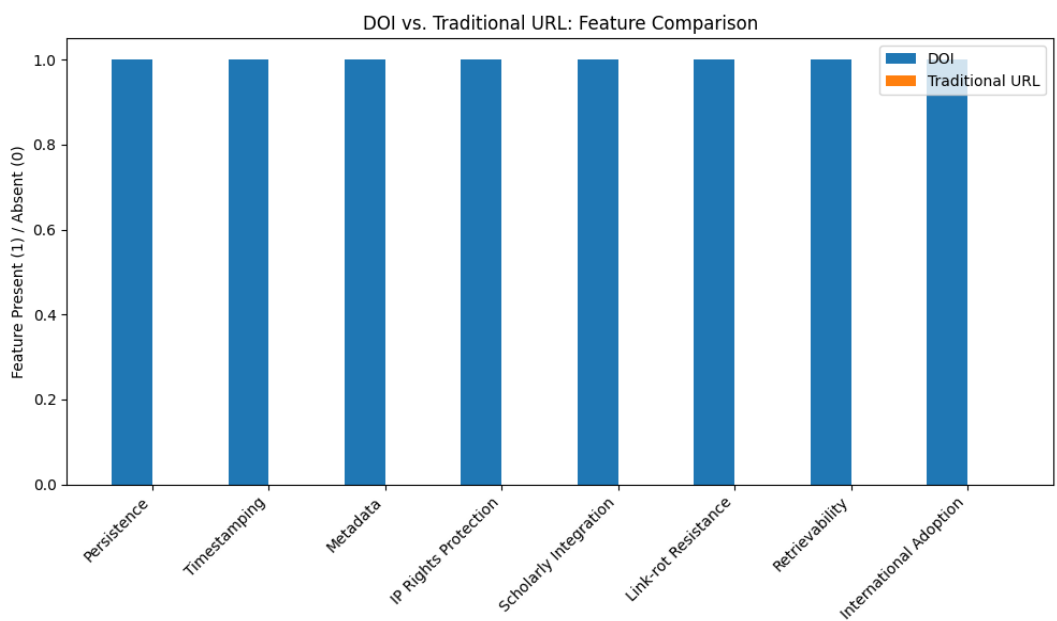

The shape (2) of a number represents DOI vs. Traditional URL: Feature

5. **Its Role in Research Credibility and Intellectual Property**

The DOI (Digital Object Identifier) enhances the credibility of scientific research by providing a permanent and unique link to scholarly publications. It ensures accurate documentation of original sources and facilitates their verification, thereby reducing academic plagiarism. DOIs are associated with peer-reviewed research that has undergone rigorous scientific evaluation, and they provide comprehensive metadata about authors, institutions, and publication dates. DOIs also assist in measuring the impact of research through citation tracking and enable linking research to supporting data, making result verification possible. This system contributes to distinguishing reputable journals from predatory ones and supports the construction of a reliable citation network that fosters the transparent accumulation of scientific knowledge.

- Permanent and Reliable Identification: DOI provides a stable method for identifying digital objects (Krauskopf & Salgado, 2023). Unlike URLs, which may change over time, the DOI remains unchanged, ensuring the link to the object remains functional even if its web location changes (Editorial, 2017). This guarantees continuity in information access (Celko, 2010).

- Facilitating Discovery and Access: DOI serves as a system that directs users directly to the desired digital object. By clicking on a DOI link, users can access the object or information about it, even if it has been moved or updated (Liu, 2021).

- Enabling Accurate Citation: DOI allows researchers and authors to cite digital objects accurately and reliably. It provides standardized bibliographic information, making it easy to track and verify sources.

- Managing Intellectual Property Rights: DOI can be used to manage intellectual property rights of digital objects (Paskin, 1999). It can be linked to information about copyright and licensing, helping to protect the rights of authors and publishers (Lizong et al., 2010).

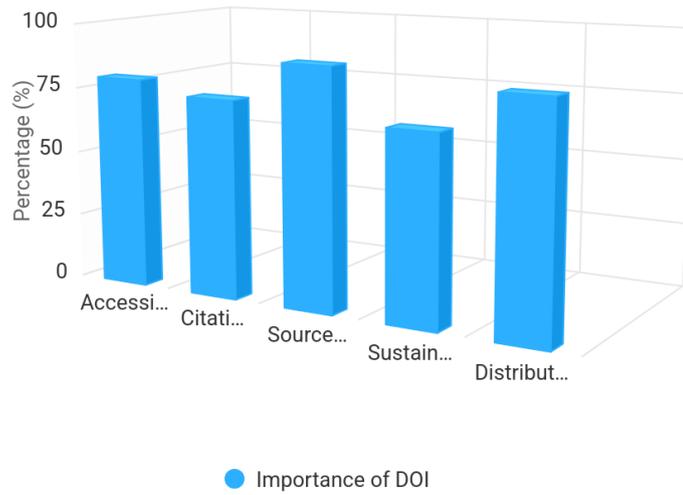

The shape (3) Percentage Importance of DOI in Different Aspects of Scientific

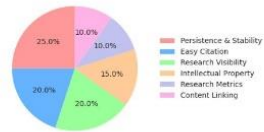
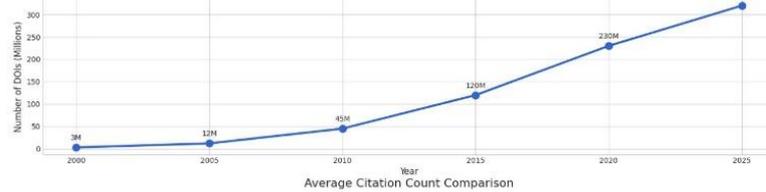
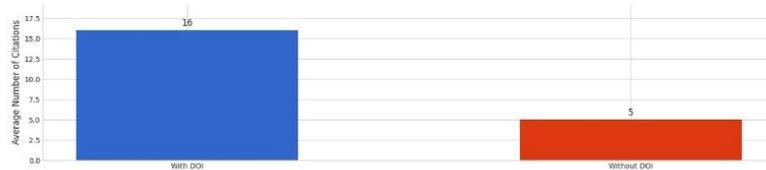

The shape (4) of a number represents Benefits of DOI in Scientific Research

The chart shows the percentages of the main benefits associated with using DOI:

- Persistence & Stability: 25%
    - This is the greatest benefit, showing that DOI helps maintain permanent access to scientific content.
- Easy Citation: 20%
- Research Visibility: 20%
- Intellectual Property: 15%
- Research Metrics: 10%
- Content Linking: 10%

Growth in DOI Usage – Line Chart in the Middle This chart illustrates the growth in the number of DOI identifiers used from the year 2000 to the projected figure in 2025:

- 2000: 3 million
- 2005: 12 million
- 2010: 45 million
- 2015: 120 million
- 2020: 230 million
- 2025 (estimated): 320 million

The data shows rapid growth, indicating increasing global reliance on DOIs in research and publications.

Average Citation Comparison – Bar Chart at the Bottom The chart compares the average number of citations for articles with a DOI and those without:

- Articles with DOI: 16 citations on average
- Articles without DOI: Only 5 citations

This highlights the importance of DOIs in enhancing research visibility and increasing its use in academic circles.

6. **Analysis of Crossref Official Reports for the Year 2025**

   Below is a detailed analysis of Crossref data as of May 6, 2025, followed by an organized table of information. (Crossref, 2025)

Analysis:

1. Overall Scope:

- Total records in the database: 170,003,248

- Number of publishers (Prefixes): 30,519

- These figures reflect the significant expansion in DOI usage across various publishing sources.

2. Publications by Type:

- Number of journals: 147,128

    o Of these, journal DOIs: 115,482,960 → represent the vast majority.

- Number of conferences: 125,878

    o Conference DOIs: 9,141,535

- Number of databases: 67,717

    o Database DOIs: 3,155,440

- Number of standards: 492,497

    o Standards DOIs: 401,150

- Books: Not Available (N/A)

3. Special DOI Types:

- DOIs for research grants: 163,195

- DOIs for peer reviews: 750,385

- Published components (such as figures or tables): 8,793,471

4. Links and Citations:

- "Cited-by" links: More than 1.8 billion

- However, the number of articles with uploaded references is 0, which is strange or may indicate a lack of reference submission.

5. By Title Level:

- Journal titles: 72,769
- Book titles: 2,239,584
- Conference titles: 75,469
- Theses: 798,930
- Reports: 827,643
- Standards: 400,585
- Databases: 50,219

6. Funding and Transparency:

- DOIs containing funding data (FundRef): 2,947,660 → reflects increasing transparency in research funding.

Table number (2) represents Statistical Summary of Crossref Data for the Year 2025

| Category | Count |
|---|---|
| Total records in the database | 170,003,248 |
| Number of prefixes (Prefixes) | 30,519 |
| Number of journals | 147,128 |
| - DOIs for journals | 115,482,960 |
| Number of books | Not Available |
| - DOIs for books | Not Available |
| Number of conferences | 125,878 |
| - DOIs for conferences | 9,141,535 |
| Published components | 8,793,471 |
| Number of standards | 492,497 |
| - DOIs for standards | 401,150 |
| Number of databases | 67,717 |
| - DOIs for databases | 3,155,440 |
| DOIs for research grants | 163,195 |
| DOIs for peer review | 750,385 |
| DOIs containing funding data | 2,947,660 |
| Number of "Cited-by" links | 1,804,719,872 |
| DOIs with a single citation | Under development |
| DOIs for journal titles | 72,769 |
| DOIs for book titles | 2,239,584 |
| DOIs for conference titles | 75,469 |
| DOIs for theses | 798,930 |
| DOIs for reports | 827,643 |
| DOIs for standards titles | 400,585 |
| DOIs for database titles | 50,219 |

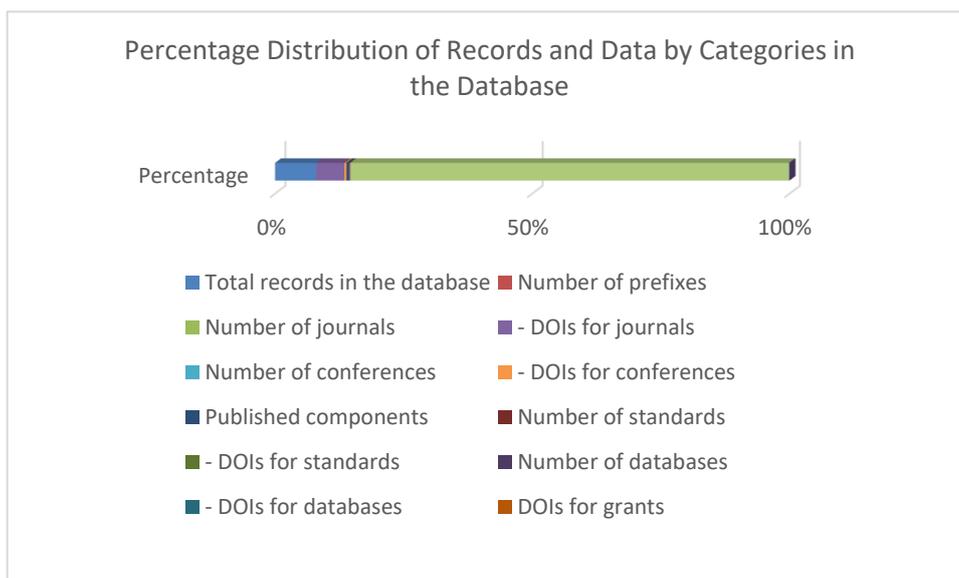

Shape number (5) Percentage Distribution of Records and Data by Categories in the Database

7. **Key Statistics on DOI Usage up to May 2025**

- Number of Registered Records: Crossref surpassed 165 million metadata records by March 2025, including journal articles, books, research papers, datasets, reports, blogs, and more.

- Number of Members: Over 20,000 members from 150 countries have joined Crossref, including publishers, libraries, research institutions, and funding bodies. (Wikipedia)

- Number of DOI Resolutions: Around 1.1 billion DOI resolution attempts are performed monthly, reflecting the widespread use of digital identifiers in research and referencing.

- Data File Size: The public data file for 2025 is approximately 197 GB, containing all Crossref records up to March 2025.

Distribution of Registered Content Types

According to Crossref data, the registered content using DOIs is distributed as follows:

- Journals and Articles: Approximately 71% of records. (press, n.d.)

- Books: Around 17%.

- Conference Papers: About 6%.

- Other Content: Includes datasets, reports, blogs, and more.

Global Expansion of DOI Usage

Reports indicate that Crossref has seen a 512% increase in the number of abstracts included in metadata since 2018, along with a 3004% growth in linking between preprints and final published articles.

Access to Public Data

Crossref offers a free annual public data file, which can be downloaded via Academic Torrents or Amazon S3. The file contains records in JSON-lines or SQLite formats, making it easy to use for analysis and research. (Crossref Community Forum, www-crossref-org.ezproxy.csu.edu.au).

The above statistics show a set of data that can be analyzed as follows:

1- Number of recorded entries

- Count: Crossref has surpassed 165 million records.

- Analysis: This number indicates the large volume of data being registered using DOI (Digital Object Identifier). These records include journal articles, books, research papers, scientific data, and reports, reflecting the massive growth in the use of DOI to document and provide access to diverse scientific and research content worldwide. This growth reflects an increase in scientific production and the role of DOI in ensuring permanent access to research.

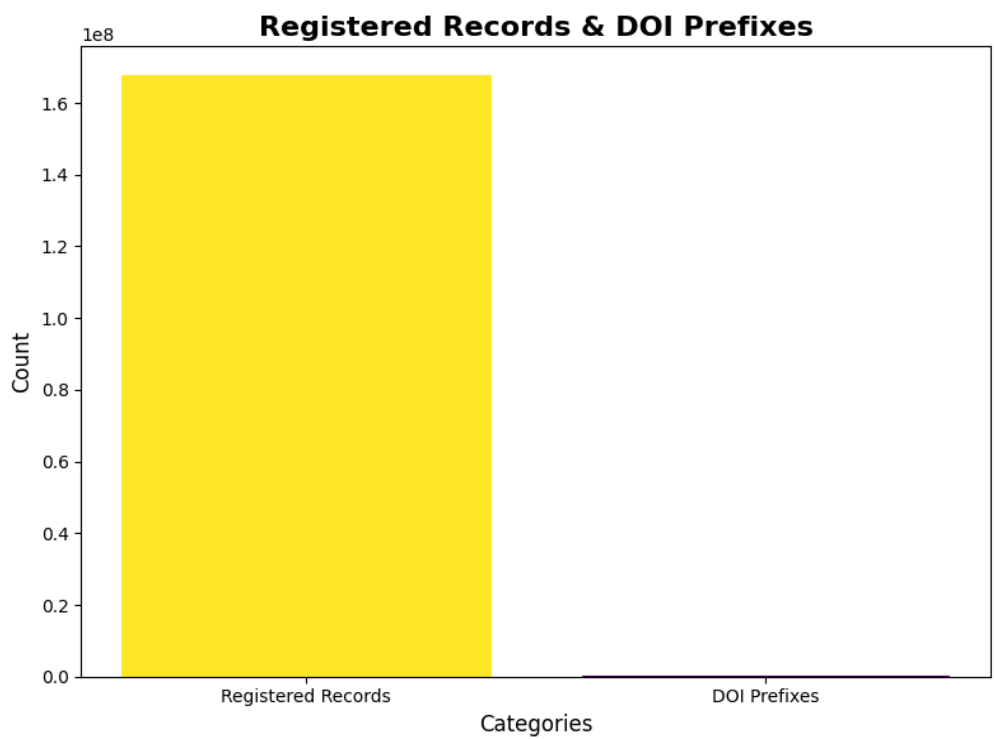

Shape number (6) Number of recorded entries

2- Number of members

- Count: Over 20,000 members from 150 countries.

- Analysis: Crossref has seen a significant increase in the number of members collaborating to provide DOIs. These members include publishers, research institutions, libraries, and funding agencies that use DOIs to register and document their publications. This growth reflects the global reliance of research institutions on DOIs as a reliable and standardized way to represent scientific content and enhance its accessibility, contributing to strengthening global scientific collaboration.

3- Number of DOI Resolutions

- Count: 1.1 billion resolution attempts monthly.

- Analysis: The high number indicates the widespread use of DOIs in research. When a specific DOI is requested, the user is "directed" to the original source of the content (such as a scientific article or book) via a stable link. These resolutions indicate that researchers and readers trust DOIs to access scientific content reliably. This number directly reflects the extent of DOI usage and its prevalence in the academic community.

4- Public Data Size

- Size: 197 GB.

- Analysis: The public file provided by Crossref contains massive data (197 GB). This data includes scientific records for all registered DOIs over the years. Crossref offers this data to researchers and software developers for analysis and research, contributing to improving transparency in accessing scientific research. The large data size reflects the diversity and continuous expansion in the use of DOIs.

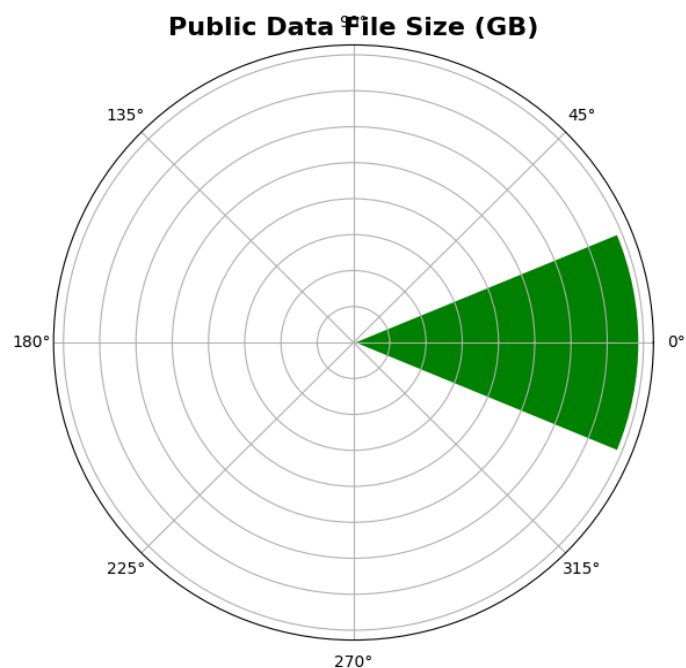

Shape number (7) Public Data Size

5- Distribution of Content Types Registered with DOI

- Journals and Articles: 71%

- Books: 17%

- Conference Papers: 6%

- Other Content: 6%

- Analysis: Research published in journals and articles represents the largest portion of content registered with DOI, reflecting the widespread adoption of DOI as a primary tool for documenting academic research across various fields. Books make up only 17%, which may suggest that research published in the form of shorter articles is more common. Conference papers and other content include preliminary research and reports that benefit from DOI to improve their accessibility.

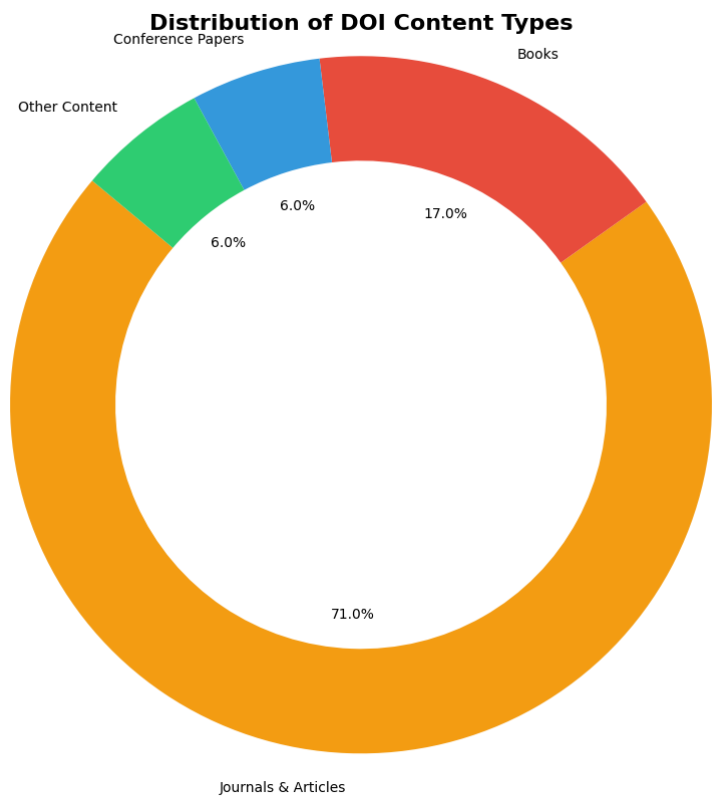

Shape number (8) Distribution of Content Types

6- Global Expansion in DOI Usage

- Increase in Listed Records: 512%

- Increase in Links Between Preprints and Final Articles: 3004%

- Analysis: The significant increase in the number of listed records indicates that more research papers and articles are now being registered using DOI. This reflects a development in how scientific information is organized and documented institutionally. The large increase in links between preprints (such as research preprints) and final articles means there is a growing trend in documenting and connecting preliminary and final research, improving access to research and encouraging the publication of initial research findings for review and critique.

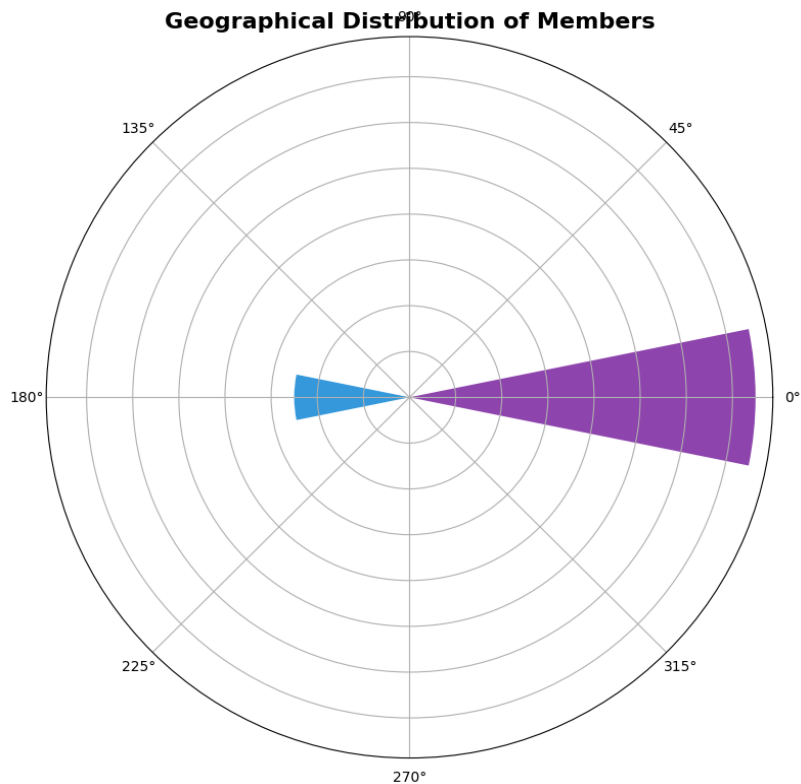

Shape number (9) Geographical Distribution of Members

The data shows that DOI has become a vital tool in the global scientific system, enhancing access, transparency, and accurate documentation of research. The increase in the number of records, members, and resolution attempts demonstrates the academic community's growing adoption of this tool, making it essential in steering scientific research towards a more integrated and inclusive future.

**Conclusion**

In addition to the points addressed, it should be noted that the use of the Digital Object Identifier (DOI) plays a crucial role in enhancing the research experience for both readers and researchers. It facilitates quick access to the full texts of research papers, thereby increasing the opportunities for reviewing and utilizing information effectively in future studies. DOI is also an effective tool in combating information overload, as readers can rely on the trusted links it provides, ensuring that they access the correct content without confusion.

Moreover, DOI enhances the value of research in both academic and commercial contexts, as its presence is an attractive factor for funders and sponsors, contributing to further investments in scientific research. A complex network of citations and links between different studies is created through DOI, enabling researchers to form new contexts and conclusions supported by strong evidence.

Furthermore, DOI strengthens the legal security of research by providing documented proof of intellectual property ownership, which bolsters the rights of authors and researchers. In light of increasing copyright laws and regulations, DOI serves as an important guarantee for protecting creators' rights, thus encouraging more innovation and creativity in the scientific community.

Ultimately, it is clear that the Digital Object Identifier (DOI) is not merely a technical tool but an essential component in building a reliable research system that reflects the efforts made to achieve high quality and trustworthy results in various fields of scientific research. Therefore, academic institutions and researchers should promote the use of DOI and intensify efforts to raise awareness of its importance, in order to ensure the highest standards of integrity and transparency in scientific research and to foster confidence in its results among the scientific community and the wider public.